\documentclass[11pt]{article}

\usepackage{amsmath,amssymb,graphicx,booktabs}
\usepackage{subcaption}
\usepackage{placeins}
\usepackage[margin=1in]{geometry}
\usepackage{lmodern}
\usepackage{xcolor}
\usepackage{textcomp} 
\usepackage{needspace}
\usepackage{etoolbox}
\AtBeginEnvironment{abstract}{%
  \setlength{\parindent}{10pt}%
  \setlength{\parskip}{6pt}%
  \setlength{\leftskip}{0pt}%
  \setlength{\rightskip}{0pt}%
}
\makeatletter
\renewcommand{\maketitle}{%
  \begin{center}
    {\LARGE \@title \par}
    \vskip 4em
    {\large \lineskip .8em \begin{tabular}[t]{c}\@author\end{tabular}\par}
    \vskip .8em
    {\small  \@date \par} 
    \vskip 1.5em
  \end{center}
}
\makeatother
\definecolor{linkblue}{HTML}{0070C0}
\usepackage{xurl} 
\usepackage{hyperref}
\hypersetup{
  colorlinks=true,
  urlcolor=linkblue,
  linkcolor=linkblue,
  citecolor=black
}

\title {Wigner's Friend as a Circuit: Inter-Branch Communication Witness Benchmarks on Superconducting Quantum Hardware}
\author{Christopher Altman\,{\raisebox{0.6ex}{\scriptsize *}}}
\date{January 2026}

\newlength{\footstarindent}

\begin{document}
\maketitle

\makeatletter
\let\origfootnoterule\footnoterule
\renewcommand{\footnoterule}{%
  \vspace*{10pt}%
  \hrule width 0.2\columnwidth
  \vspace*{10pt}%
}
\makeatother

\begin{abstract}
\noindent We implement and benchmark on IBM Quantum hardware the circuit family proposed by Violaris (2026) for estimating operational ``inter-branch communication'' witnesses, i.e., correlations in classical measurement records produced by compiled Wigner's-friend-style circuits. Concretely, we realize a five-qubit instance of the protocol as an inter-register message-transfer \emph{pattern} in the circuit (not physical signaling), and evaluate its behavior under realistic device noise and compilation constraints. The circuit encodes a branch-conditioned evolution in which an observer subsystem $F$ (``Wigner's friend'') evolves differently depending on the state of a control qubit $Q$, followed by a controlled transfer operation that probes correlations between branch-conditioned histories (conditional measurement contexts within a single compiled circuit). We evaluate coherence-sensitive witnesses on a specified four-qubit subset.
Executing on the \texttt{ibm\_fez} backend ($N=20{,}000$ shots), we obtain population-based visibility $V = 0.8771 \pm 0.0034$, coherence witnesses $W_X = 0.8398 \pm 0.0038$ and $W_Y = -0.8107 \pm 0.0041$, and phase-sensitive magnitude $C_\mathrm{mag} = (W_X^2 + W_Y^2)^{1/2} = 1.1673 \pm 0.0040$.

\vspace{8pt}
\noindent This work does not claim to test or discriminate among interpretations of quantum mechanics. Unitary interpretations that reproduce standard quantum predictions remain operationally equivalent for these circuits. Instead, we frame the experiment as a publicly reproducible constraint pipeline: given a parameterized family of non-ideal channels (e.g., dephasing or collapse-like perturbations) inserted at specified circuit locations, the coherence-sensitive witnesses define detectability thresholds relative to calibrated device noise. The accompanying reproducibility bundle provides full provenance (job IDs, calibration snapshots, software versions) enabling independent verification and extension.
\end{abstract}

\begingroup
  \makeatletter
  \renewcommand{\footnoterule}{%
    \vspace*{10pt}
    \hrule width 0.4\columnwidth
    \vspace*{10pt}
  }
  \setlength{\footnotesep}{8pt}
  \renewcommand\@makefntext[1]{\noindent #1}
  \makeatother

  \renewcommand\thefootnote{}%
  
  \footnotetext{%
    {\scriptsize\sffamily\linespread{0.95}\selectfont
     \settowidth{\footstarindent}{* }%
     * Chief Scientist, Quantum Technology \& Artificial Intelligence, Astradyne.\;\\
     \leavevmode\hspace*{\footstarindent}%
     Web: \href{https://lab.christopheraltman.com}{lab.christopheraltman.com}\,\textperiodcentered\,Email: \href{mailto:x@christopheraltman.com}{x@christopheraltman.com}.%
    }%
    
    \makeatletter
\let\footnoterule\origfootnoterule
\makeatother
    
  }%
  \addtocounter{footnote}{-1}%
\endgroup

\section{Introduction}
\label{sec:intro}

The Wigner's friend thought experiment~\cite{wigner1961remarks} and its extensions~\cite{frauchiger2018quantum,brukner2018no} probe foundational questions about unitary quantum dynamics, measurement, and observer perspectives. The quantum information community has increasingly adopted a ``circuits-as-foundations'' approach: encoding these thought experiments as explicit quantum circuits that can be analyzed, simulated, and executed on real quantum hardware~\cite{proietti2019experimental,bong2020strong}. 
\Needspace{6\baselineskip}

 Violaris~\cite{violaris2026interbranch} proposes a family of inter-branch communication protocols in which operations on a superposed ``friend'' register enable conditional message transfer between branch-conditioned internal states. Branch interference effects, if present, should manifest in specific correlations that can be probed via appropriately chosen observables. The protocol provides a concrete circuit template suitable for near-term quantum processors.

\paragraph{Contributions of this work.}
We provide a publicly reproducible IBM-hardware execution of a minimal branch-transfer circuit benchmark with the following features:
\begin{enumerate}
    \item \textbf{Coherence-witness diagnostics:} Beyond the standard population-based visibility $V$, we measure multi-qubit Pauli-parity correlators ($W_X$, $W_Y$) that are sensitive to off-diagonal coherences potentially missed by $V$.
    \item \textbf{Backend-matched noise modeling:} We compare hardware results against Qiskit Aer simulations using noise models derived from contemporaneous \texttt{ibm\_fez} calibration data~\cite{qiskitaer}.
    \item \textbf{Nonunitary channel constraint pipeline:} We demonstrate how parameterized dephasing channels can be constrained relative to observed witness values and device-noise bands.
    \item \textbf{Full provenance:} Job IDs, calibration snapshots, software versions, and SHA256 hashes for all artifacts enable independent verification.
\end{enumerate}

We emphasize what this work does \emph{not} claim: it does not uniquely test or confirm any interpretation of quantum mechanics. The Many-Worlds interpretation, Copenhagen-style collapse models, and other unitary completions all predict identical measurement statistics for these circuits when the evolution is unitary. The experiment's value lies in (a) demonstrating that branch-transfer primitives behave consistently with unitary predictions on real hardware, and (b) establishing a methodology for constraining specific nonunitary perturbations should future experiments seek to test such models.

\paragraph{Related work.}
Experimental implementations of Wigner's-friend scenarios on quantum processors include the photonic realization by Proietti et al.~\cite{proietti2019experimental} and the extended Wigner's friend test by Bong et al.~\cite{bong2020strong}. Those works focus on Bell-type inequalities and local friendliness assumptions. Our work differs in targeting the inter-branch communication primitive specifically and in emphasizing coherence-witness diagnostics as a complement to population-based metrics. The IBM Quantum platform~\cite{ibmquantum} and Qiskit software stack~\cite{qiskit,qiskitibmruntime} provide the infrastructure for our hardware execution.

\paragraph{Related constraint results.}
Mukherjee and Hance~\cite{mukherjee2025timelike} analyze timelike constraints on ``absoluteness of observed events'' assumptions in Wigner's-friend settings. Our work is complementary: we focus on an implementable circuit primitive inspired by inter-branch communication protocols and report its empirical compilation and noise behavior on superconducting quantum hardware, without adopting observer-independent event absolutism.

\section{Protocol and Metrics}
\label{sec:protocol}

\begin{figure}[!t]
  \centering
  \IfFileExists{branch_transfer_circuit.png}{
    \includegraphics[width=0.92\linewidth]{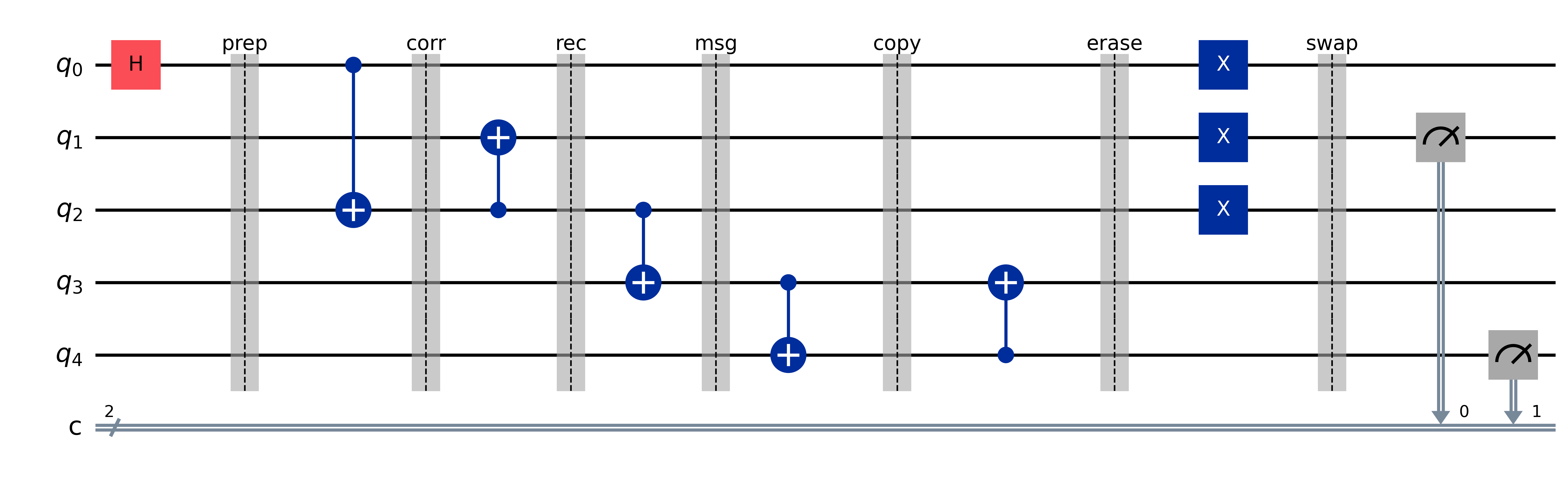}
  }{
    \fbox{\parbox{0.92\linewidth}{\centering
      \textbf{Circuit diagram placeholder}\\[0.25em]
      Add a schematic at \texttt{branch\_transfer\_circuit.png} (e.g., exported from Qiskit or IBM Quantum Composer).}}
  }
  \caption{Branch-transfer circuit schematic (five-qubit primitive) as transpiled and executed on hardware. Coherence-witness measurements act on the four-qubit subset $(Q,R,F,P)$; the auxiliary qubit participates in the controlled transfer but is not measured in coherence-witness mode. Qubit labels correspond to logical register assignments; physical qubit indices after transpilation are recorded in the reproducibility bundle.}
  \label{fig:branch_transfer_circuit}
\end{figure}

\subsection{Registers and Qubit Mapping}
\label{subsec:registers}

The circuit uses five qubits in total, with four qubits measured for the coherence-witness protocol:
\begin{itemize}
    \item $Q$: Control qubit determining branch superposition (logical index 0)
    \item $R$: Reference/ancilla qubit for message encoding (logical index 1)
    \item $F$: Friend register encoding the observer's internal state (logical index 2)
    \item $P$: Probe qubit for readout correlations (logical index 3)
    \item Logical index 4: Auxiliary qubit used in the branch-transfer primitive (not measured in coherence-witness mode)
\end{itemize}

\noindent\textbf{Note on qubit labeling:} The indices 0--4 above refer to \emph{logical} register assignments in the abstract circuit. After transpilation to hardware, these are mapped to physical qubit indices on the device; the mapping is recorded in the reproducibility bundle (\texttt{assets/bundles/wigner-friend-v2b/appendix/}).

The coherence witnesses $W_X$ and $W_Y$ are four-qubit Pauli correlators measured on $(Q, R, F, P)$. The visibility protocol measures only two qubits ($R$ and $P$).

The circuit consists of three stages: (i) preparation of branch-conditioned evolution where $F$ evolves differently depending on the state of $Q$; (ii) application of a controlled message-transfer primitive that creates correlations between $R$, $F$, and $P$; and (iii) measurement in either the computational basis (for visibility) or rotated bases (for coherence witnesses).

\subsection{Visibility \texorpdfstring{$V$}{V}: A Population-Based Diagnostic}
\label{subsec:visibility}

The visibility metric $V$ (corresponding to bundle mode \texttt{rp\_z}) is extracted from Z-basis measurement populations:
\begin{equation}
V = \bigl| P(R=0 \,|\, P=1) - P(R=1 \,|\, P=1) \bigr|,
\end{equation}
where $P(R=r \,|\, P=p)$ denotes the conditional probability of measuring qubit $R$ in state $r$ given that qubit $P$ is measured in state $p$.

\textbf{Key limitation:} Because $V$ depends only on diagonal (computational-basis) populations, it can fail to detect suppression of off-diagonal density matrix elements. A dephasing channel inserted at certain circuit locations may leave $V$ nearly unchanged while destroying coherences probed by other observables. This limitation motivates direct coherence measurements.

\subsection{Coherence Witnesses
  \texorpdfstring{$W_X$}{WX},
  \texorpdfstring{$W_Y$}{WY},
  and \texorpdfstring{$C_\mathrm{mag}$}{Cmag}}
\label{subsec:witnesses}

We measure four-qubit Pauli-parity correlators on the subset $(Q, R, F, P)$:
\begin{equation}
W_X = \langle X_Q \otimes X_R \otimes X_F \otimes X_P \rangle, \qquad
W_Y = \langle Y_Q \otimes Y_R \otimes Y_F \otimes Y_P \rangle,
\end{equation}
where $X$ and $Y$ are the standard Pauli operators.

\paragraph{Basis rotation procedure.}
To measure $W_X$, we apply a Hadamard gate $H$ to each of $(Q, R, F, P)$ immediately before Z-basis measurement. The expectation value is then computed from bitstring counts as:
\begin{equation}
W_X = \frac{N_\mathrm{even} - N_\mathrm{odd}}{N_\mathrm{even} + N_\mathrm{odd}},
\end{equation}
where $N_\mathrm{even}$ ($N_\mathrm{odd}$) is the number of shots with even (odd) total parity across the four measured bits.

To measure $W_Y$, we apply $S^\dagger H$ (where $S^\dagger$ is the inverse phase gate) to each qubit before Z-measurement, then compute parity analogously.

\paragraph{Phase-sensitive magnitude.}
We define:
\begin{equation}
C_\mathrm{mag} = \sqrt{W_X^2 + W_Y^2}.
\end{equation}

\noindent\textbf{Interpretation of $C_\mathrm{mag}$:} This quantity is \emph{not} bounded by 1 and should not be interpreted as a probability or ``coherence fraction.'' Since $W_X, W_Y \in [-1, 1]$ for any quantum state, $C_\mathrm{mag}$ is bounded above by $\sqrt{2}$. For ideal statevector evolution with this circuit, $W_X^\mathrm{ideal} = 1$ and $W_Y^\mathrm{ideal} = -1$, yielding $C_\mathrm{mag}^\mathrm{ideal} = \sqrt{2} \approx 1.414$.

\section{Implementation on IBM Quantum Hardware}
\label{sec:implementation}

\subsection{Execution Parameters}

Circuits were executed on backend \texttt{ibm\_fez} via the IBM Quantum open-access plan using Qiskit IBM Runtime~\cite{qiskitibmruntime}. Key parameters:
\begin{itemize}
    \item \textbf{Shots:} 20,000 per circuit
    \item \textbf{Transpiler optimization level:} 2 for hardware runs; 1 for backend-matched noisy simulations
    \item \textbf{Measurement modes:} \texttt{coherence\_witness} (X and Y bases) and \texttt{rp\_z} (visibility)
\end{itemize}

\begin{table}[!b]
\centering
\caption{Headline metrics for the branch-transfer circuit primitive. Reported uncertainties are shot-noise derived (stored as \texttt{*\_error} fields in JSON artifacts). $C_\mathrm{mag} = \sqrt{W_X^2 + W_Y^2}$ is a phase-sensitive coherence magnitude bounded by $\sqrt{2}$ (not by 1); ideal value is $\sqrt{2} \approx 1.414$.}
\label{tab:headline}
\begin{tabular}{lccc}
\toprule
Metric & Ideal (statevector) & Backend-matched noisy sim & Hardware (\texttt{ibm\_fez}) \\
\midrule
$V$ & 1.0000 & 0.9381 & $0.8771 \pm 0.0034$ \\
$W_X$ & 1.0000 & $0.8984 \pm 0.0031$ & $0.8398 \pm 0.0038$ \\
$W_Y$ & $-1.0000$ & $-0.8972 \pm 0.0031$ & $-0.8107 \pm 0.0041$ \\
$C_\mathrm{mag}$ & 1.4142 & 1.2697 & $1.1673 \pm 0.0040$ \\
\bottomrule
\end{tabular}
\end{table}

\subsection{Provenance Box}
\label{subsec:provenance}

\begin{center}
\fbox{\parbox{0.9\textwidth}{
\textbf{Hardware Provenance} \\[0.5em]
\begin{tabular}{@{}ll@{}}
Backend: & \texttt{ibm\_fez} \\
Instance: & open-instance (open plan) \\
Shots: & 20,000 per circuit \\
Optimization level (hardware): & 2 \\
Optimization level (simulator): & 1 \\
\end{tabular}

\vspace{0.5em}
\textbf{Job IDs} \\[0.3em]
\begin{tabular}{@{}ll@{}}
Coherence X: & \texttt{d5lobdt9j2ac739k1a0g} \\
Coherence Y: & \texttt{d5locdhh2mqc739a2ubg} \\
Visibility (rp\_z): & \texttt{d5locnd9j2ac739k1b80} \\
\end{tabular}

\vspace{0.5em}
\textbf{Software Versions} \\[0.3em]
\begin{tabular}{@{}ll@{}}
Python: & 3.10.19 \\
Qiskit: & 2.3.0~\cite{qiskit} \\
Qiskit Aer: & 0.17.2~\cite{qiskitaer} \\
qiskit-ibm-runtime: & 0.45.0~\cite{qiskitibmruntime} \\
\end{tabular}

\vspace{0.5em}
\textbf{Bundle:} v2b --- SHA256 hashes for all files in \texttt{assets/bundles/wigner-friend-v2b/MANIFEST.json}
}}
\end{center}

\subsection{Backend-Matched Noise Modeling}
\label{subsec:noise}

We include baseline simulations using Qiskit Aer~\cite{qiskitaer} with noise models constructed via \\ \path|NoiseModel.from_backend()| using contemporaneous calibration snapshots of \texttt{ibm\_fez}. 
\vspace{0.5em}

These snapshots (stored in \texttt{assets/bundles/wigner-friend-v2b/appendix/}) capture gate error rates, readout errors, and $T_1$/$T_2$ times at the time of execution.
\vspace{0.5em}

Such backend-matched models are approximate: they may fail to fully capture drift, crosstalk, leakage, and coherent errors not characterized by the calibration data. We therefore treat the noisy simulation as a \emph{calibrated proxy} rather than a strict prediction, useful for understanding the noise floor but not as ground truth.

\section{Results}

Headline metrics for the branch-transfer primitive are summarized in Table~\ref{tab:headline}.

\label{sec:results}

Numeric values are extracted from bundled JSON artifacts or computed from stored fields using declared formulas (e.g., $C_\mathrm{mag} = \sqrt{W_X^2 + W_Y^2}$). All source artifacts are integrity-checkable via SHA256 hashes in \texttt{assets/bundles/wigner-friend-v2b/MANIFEST.json}.

\paragraph{Interpretation of results.}
Hardware values show degradation relative to ideal predictions, consistent with device noise. The visibility $V = 0.8771$ on hardware is lower than the backend-matched noisy simulation prediction ($V = 0.9381$), suggesting calibration drift, additional noise sources not captured by the model, or transpilation effects at optimization level 2. Coherence witnesses $W_X$ and $W_Y$ similarly show hardware degradation beyond the noisy simulation baseline.

Crucially, all measured values remain consistent with \emph{unitary evolution plus device noise}---no anomalous deviations suggesting nonunitary physics are observed. The results establish a baseline for future comparisons.

\begin{figure}[!htbp]
  \centering

  \begin{subfigure}[t]{0.49\linewidth}
    \centering
    \IfFileExists{coherence_comparison.png}{
      \includegraphics[width=\linewidth,height=0.32\textheight,keepaspectratio]{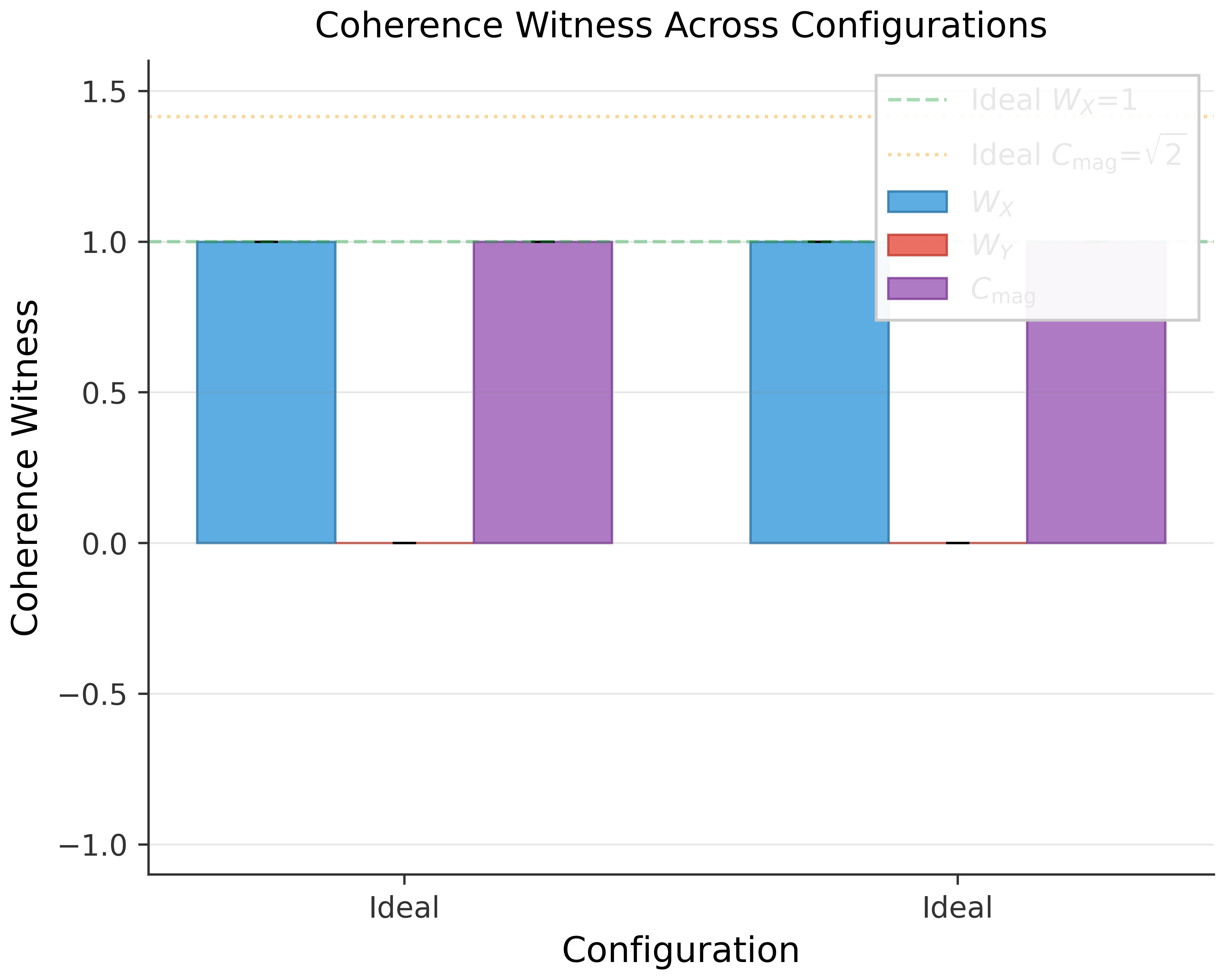}
    }{
      \fbox{\parbox{0.92\linewidth}{\centering Missing: \texttt{coherence\_comparison.png}}}
    }
    \caption{Coherence witnesses ($W_X, W_Y, C_{\mathrm{mag}}$).}
    \label{fig:coherence_comparison}
  \end{subfigure}\hfill
  \begin{subfigure}[t]{0.49\linewidth}
    \centering
    \IfFileExists{visibility_comparison.png}{
      \includegraphics[width=\linewidth,height=0.32\textheight,keepaspectratio]{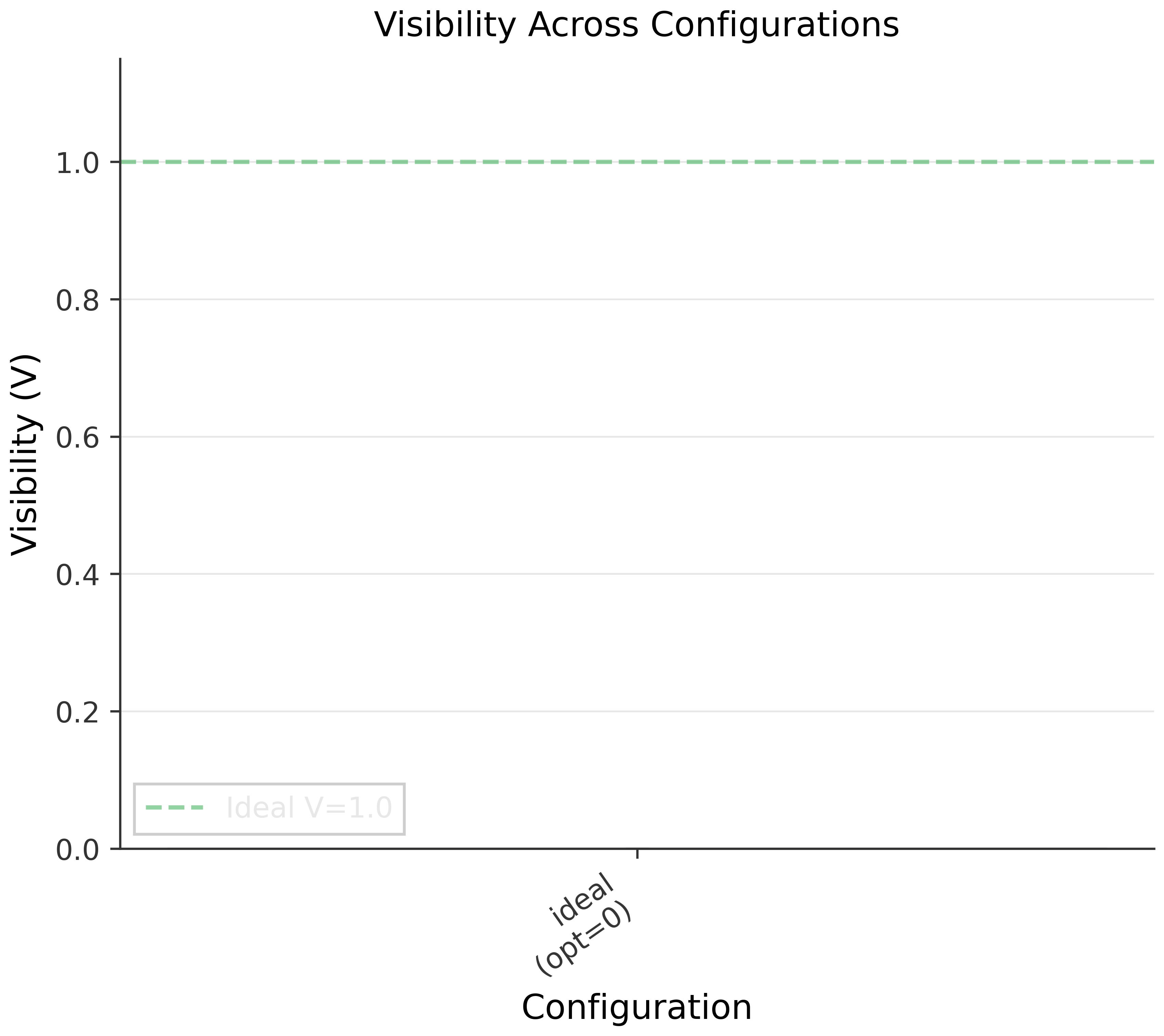}
    }{
      \fbox{\parbox{0.92\linewidth}{\centering Missing: \texttt{visibility\_comparison.png}}}
    }
    \caption{Population visibility $V$.}
    \label{fig:visibility_comparison}
  \end{subfigure}

  \caption{Primary hardware vs.\ simulation comparisons: (a) coherence witnesses across ideal, backend-matched noisy simulation, and hardware; (b) population visibility across the same conditions. Provenance and SHA256 hashes are recorded in \texttt{assets/bundles/wigner-friend-v2b/MANIFEST.json}.}
  \label{fig:primary_comparisons}
\end{figure}

\section{Nonunitary Channel Constraint Pipeline}
\label{sec:collapse}

We turn witness measurements into an exclusion bound on a parameterized channel family inserted at specified circuit locations.

\begin{enumerate}
    \item \textbf{Channel family:} Define a parameterized family of channels $\mathcal{E}_\lambda$ (e.g., single-qubit dephasing with strength $\lambda \in [0,1]$) and specify the circuit insertion point.
    \item \textbf{Predicted deviation curves:} For each $\lambda$, simulate the circuit with the channel inserted and compute predicted values of $V$, $W_X$, $W_Y$, $C_\mathrm{mag}$.
    \item \textbf{Detectability threshold:} A channel strength $\lambda$ is ``detectable'' if the predicted deviation in at least one witness exceeds the combined uncertainty band (shot noise $\oplus$ device-noise proxy derived from backend-matched simulation discrepancy).
    \item \textbf{Constraint:} The observed hardware values constrain the channel strength to $\lambda < \lambda_\mathrm{max}$ where $\lambda_\mathrm{max}$ is the detectability threshold.
\end{enumerate}

\begin{center}
\fbox{\begin{minipage}{0.95\linewidth}
\textbf{Worked example (phase-flip dephasing estimate).}
Consider a single-qubit phase-flip (dephasing) channel applied to the friend register after the branch split,
\begin{equation}
\mathcal{E}_\lambda(\rho) = (1-\lambda)\rho + \lambda\, Z\rho Z .
\end{equation}
For observables containing Pauli $X$ or $Y$ on the friend qubit (which anticommute with the error operator $Z$) on the friend qubit, the channel attenuates expectations by a factor $(1-2\lambda)$. To first order, the coherence witnesses scale as
\begin{equation}
W_X(\lambda)\approx (1-2\lambda)\,W_X(0),\qquad W_Y(\lambda)\approx (1-2\lambda)\,W_Y(0).
\end{equation}
Using the ideal value $W_X(0)\approx 1$ and the measured hardware value $W_X^\mathrm{(hw)}=0.8398$, we obtain the illustrative estimate
\begin{equation}
\lambda_\mathrm{est} \approx \tfrac{1-W_X^\mathrm{(hw)}}{2} = 0.080.
\end{equation}
In the full pipeline we use backend-matched noisy simulation to define the detectability band and report the corresponding $\lambda_\mathrm{max}$ as the conservative bound for the chosen insertion point and channel family.
\end{minipage}}
\end{center}

\textbf{Terminology note:} We refer to the channel family as a ``parameterized dephasing/nonunitary channel toy model'' rather than labeling it with interpretation-specific names (e.g., ``Copenhagen collapse''). The constraint applies to the channel family under the assumed insertion point and noise calibration; it does not uniquely identify physical mechanisms.

\begin{figure}[!htbp]
  \centering

  \begin{subfigure}[t]{0.49\linewidth}
    \centering
    \includegraphics[width=\linewidth,height=0.30\textheight,keepaspectratio]{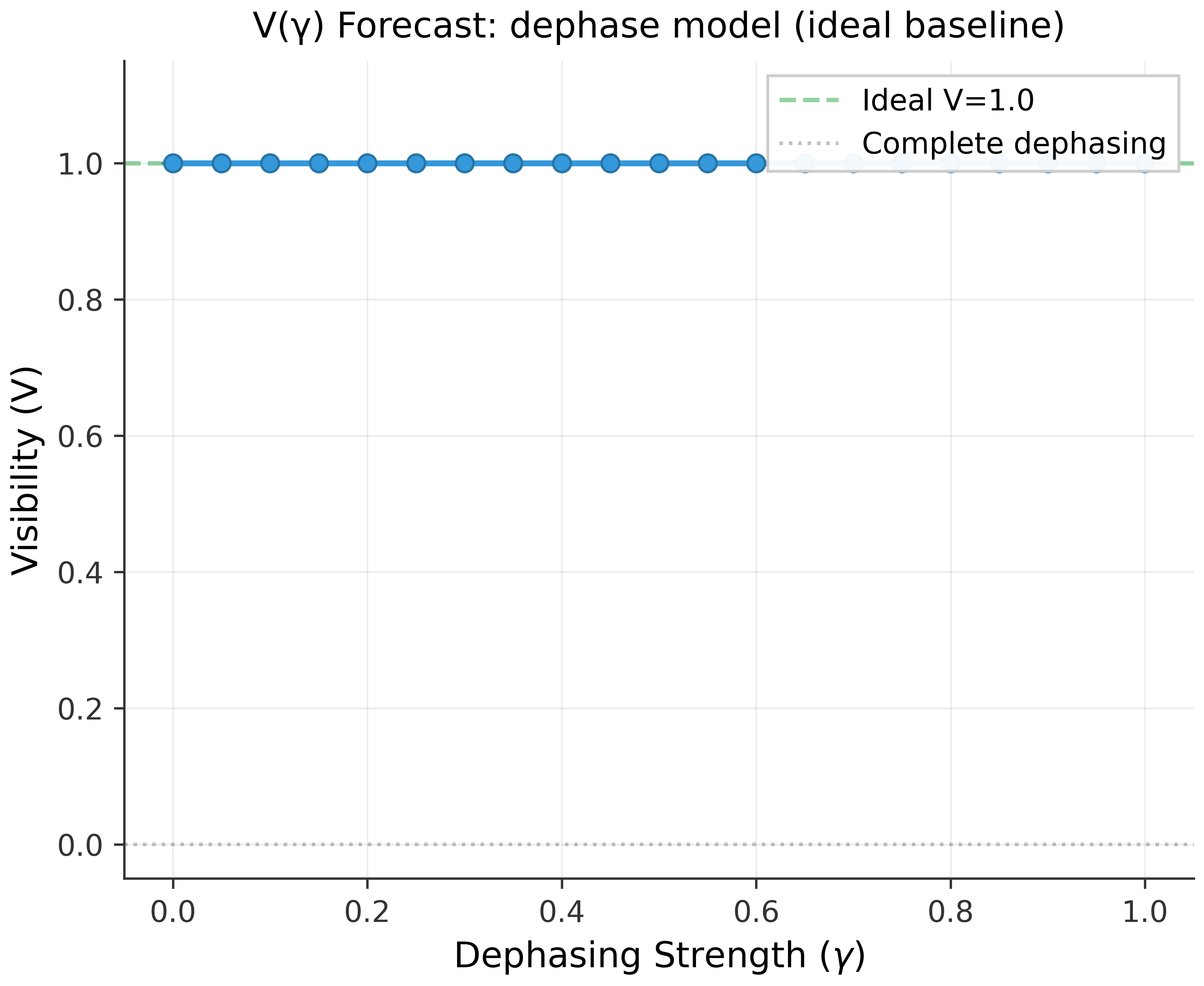}
    \caption{Ideal baseline.}
    \label{fig:collapse_forecast_ideal}
  \end{subfigure}\hfill
  \begin{subfigure}[t]{0.49\linewidth}
    \centering
    \includegraphics[width=\linewidth,height=0.30\textheight,keepaspectratio]{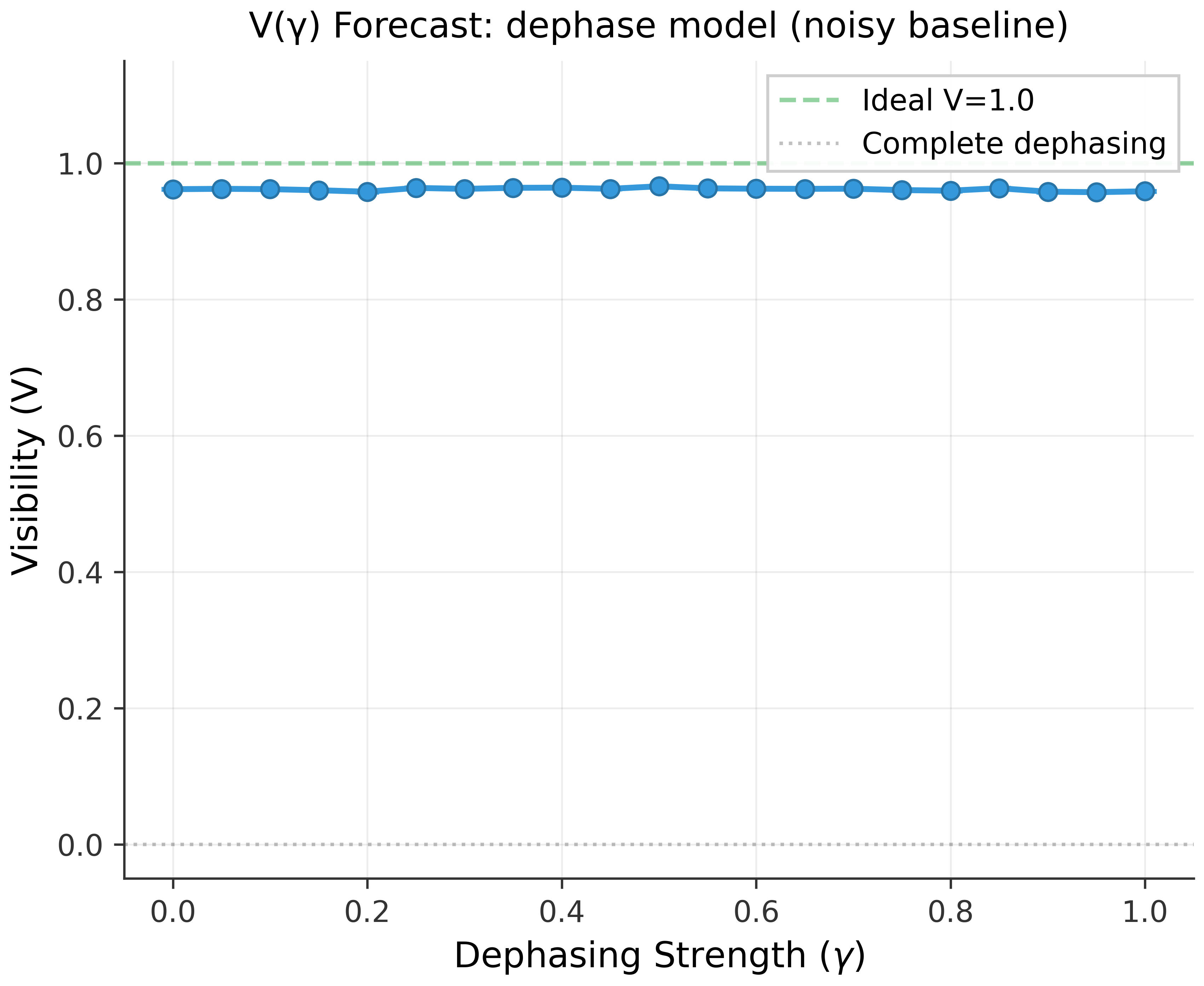}
    \caption{Noisy (backend-matched) baseline.}
    \label{fig:collapse_forecast_noisy}
  \end{subfigure}

  \caption{Dephasing-channel forecast for the branch-transfer circuit under (a) ideal and (b) backend-matched noisy conditions.}
  \label{fig:collapse_forecast_stack}
\end{figure}

The bundled forecast panels (Figures~\ref{fig:collapse_forecast_ideal} and~\ref{fig:collapse_forecast_noisy}) illustrate how coherence witnesses degrade as channel strength increases, while diagonal visibility can remain comparatively insensitive depending on where the channel is inserted. This motivates the use of $W_X$ and $W_Y$ as primary diagnostics for any future searches for nonunitary deviations.

\section{Interpretation }
\label{sec:interpretation}

\paragraph{What this does not establish.}
This experiment does not uniquely test, confirm, or refute any interpretation of quantum mechanics. The Many-Worlds interpretation, Copenhagen interpretation, relational quantum mechanics, QBism, and other frameworks that reproduce standard quantum mechanical predictions are \emph{operationally equivalent} for these circuits. A visibility of $V \approx 0.88$ and coherence magnitude $C_\mathrm{mag} \approx 1.17$ are consistent with all such interpretations plus realistic device noise.

\paragraph{What it does establish.}
The experiment demonstrates:
\begin{enumerate}
    \item The branch-transfer circuit primitive behaves consistently with unitary predictions within device noise on IBM Quantum hardware.
    \item Coherence witnesses $W_X$ and $W_Y$ provide off-diagonal sensitivity complementary to population-based visibility.
    \item A methodology exists for constraining specific nonunitary channel families relative to hardware noise floors.
\end{enumerate}

\vspace{8pt}

\paragraph{When could this become interpretation-relevant?}\mbox{}\\[0.6\baselineskip]
In principle, if future experiments observed statistically robust deviations from unitary predictions that persisted across:

\begin{itemize}
    \item multiple hardware backends with independent calibration,
    \item different transpilation and error-mitigation strategies,
    \item exhaustive exclusion of hardware artifacts (crosstalk, leakage, calibration drift, coherent errors),
    \item systematic variation of circuit parameters,
\end{itemize}
and if such deviations matched the signature of a specific nonunitary model while being inconsistent with any plausible noise mechanism, this could constitute evidence relevant to interpretation debates. The present work does not observe any such deviation; it establishes baseline methodology for potential future investigations.

\section{Conclusions and Future Work}
\label{sec:conclusion}

We have presented a reproducible IBM-hardware execution of inter-branch communication circuit primitives with coherence-witness diagnostics, backend-matched noise modeling, and a nonunitary-channel constraint pipeline. The experiment establishes baseline measurements and methodology without claiming interpretation-level conclusions.

\paragraph{Outlook.} 
This work serves as a proof-of-concept demonstration on superconducting qubits. Ongoing follow-up studies aim to replicate the same inter-branch message-transfer primitive across multiple quantum computing modalities---gate-model superconducting and ion-trap devices, neutral-atom analog platforms, photonics, and annealing-style hardware---varying connectivity and compilation constraints while preserving identical logical structure. The goal is to quantify a cross-modality ``compilation tax'' (e.g., depth/2Q-gate overhead and fidelity loss relative to an ideal logical circuit) and develop a generalized connectivity scaling law for realizing inter-branch communication primitives under realistic hardware constraints.

\paragraph{Future directions aligned with Violaris' complexity axis.}
Violaris~\cite{violaris2026interbranch} discusses how protocol complexity scales with branch divergence. Natural extensions include:
\begin{itemize}
    \item \textbf{Branch divergence scaling:} Represent friend-0 and friend-1 states as $k$-qubit bitstrings whose Hamming distance grows under branch evolution. The branch-swap operation then becomes a longer $X$-string (product of $X$ gates), yielding a complexity-vs-coherence scaling benchmark.
    \item \textbf{Multi-backend replication:} Execute on multiple IBM Quantum backends and other platforms to assess reproducibility and identify backend-specific artifacts.
    \item \textbf{Calibration-synchronized repeats:} Run experiments immediately after fresh calibration cycles to minimize drift effects.
    \item \textbf{Error mitigation:} Apply randomized compiling, Pauli twirling, or zero-noise extrapolation (ZNE) to witness observables, enabling tighter constraints on nonunitary channels.
\end{itemize}

\section{Reproducibility}
\label{sec:reproducibility}

All source code, analysis notebooks, calibration metadata, and job provenance for the results reported in this paper are publicly available in the repository:
  \vspace*{10pt}%
  
  \hspace*{30pt}%
  \href{https://github.com/christopher-altman/ibm-qml-kernel}{\nolinkurl{https://github.com/christopher-altman/ibm-qml-kernel}}
  \vspace*{10pt}%
  
  and are permanently archived on Zenodo at
\href{https://doi.org/10.5281/zenodo.18335978}{doi:10.5281/zenodo.18335978}.
  \vspace*{10pt}%

A dedicated release bundle (tag \texttt{wigner-friend-v2b}) provides the exact artifacts used to generate the figures and numerical values, including execution scripts, IBM Quantum job IDs, backend calibration snapshots, generated plots, and a SHA256 manifest for full provenance. 
The bundle can be obtained at:
  \vspace*{10pt}%

\hspace*{30pt}%
  \href{https://github.com/christopher-altman/ibm-qml-kernel/releases/tag/v1.0-wigner-branch-benchmark}{Release: v1.0-wigner-branch-benchmark}
  \vspace*{10pt}%

Independent verification requires only a standard Python environment with Qiskit~2.3.0 and the packages listed in \texttt{\texttt{requirements.txt}}.

\Needspace{12\baselineskip}

The following commands reproduce the recorded workflow:

\paragraph{Supplementary bundle contents.}
The reproducibility bundle contains additional diagnostic figures not included in the main paper for clarity: \texttt{pr\_distribution.png} (measurement outcome distributions) and \texttt{visibility\_vs\_opt\_level.png} (transpiler optimization-level sensitivity). These remain available in the bundle's \texttt{} directory for readers interested in extended diagnostics.
  \vspace*{2pt}%

\begin{verbatim}
# Verify IBM Quantum connectivity
python -c "from qiskit_ibm_runtime import QiskitRuntimeService as S; \
  s=S(); bs=s.backends(simulator=False, operational=True); \
print('n_backends=', len(bs))"

# Hardware coherence witness (X + Y bases)
python -m experiments.branch_transfer.run_ibm \
  --backend ibm_fez --mode coherence_witness \
  --include-y-basis --shots 20000 --optimization-level 2

# Hardware visibility (rp_z mode)
python -m experiments.branch_transfer.run_ibm \
  --backend ibm_fez --mode rp_z --mu 1 \
  --shots 20000 --optimization-level 2

# Backend-matched noisy simulations
python -m experiments.branch_transfer.run_sim \
  --mode coherence_witness --include-y-basis --mu 1 \
  --shots 20000 --noise-from-backend ibm_fez

python -m experiments.branch_transfer.run_sim \
  --mode rp_z --mu 1 --shots 20000 \
  --noise-from-backend ibm_fez

# Generate analysis figures
python -m experiments.branch_transfer.analyze \
  --artifacts-dir artifacts/branch_transfer \
  --figures-dir artifacts/branch_transfer/figures --plot-all
\end{verbatim}
  \vspace*{2pt}%

\section*{Acknowledgments}

The author thanks Maria Violaris for proposing the operational witness protocol family implemented in this work and for helpful correspondence on the draft; Jonte Hance and Sumit Mukherjee for comments on related temporal correlation scenarios; Anton Zeilinger for hosting two consecutive residential quantum foundations fellowships in Austria that helped shape the conceptual framing of observer scenarios; Daniel Greenberger and \v{C}aslav Brukner for stimulating discussions during that period; and Avshalom Elitzur and Paul Werbos for longstanding mentorship on time-symmetric and operational formulations of quantum theory. This work was conducted on superconducting quantum hardware through the IBM Quantum platform.


\bibliographystyle{unsrt}

\end{document}